\documentclass{article}
\topskip=\baselineskip
\let\al=\alpha
\let\bt=\beta
\let\gm=\gamma
\let\dl=\delta
\let\ep=\epsilon

\let\sg=\sigma
\let\la=\langle
\let\ra=\rangle
\let\pa=\partial
\let\e=\emph

\let\ct=\cite

\let\bv=\mathbf
\let\mr=\mathrm
\let\dt=\cdot
\let\del=\nabla
\let\dg=\dagger

\let\q=\widehat

\let\h=\hbar

\let\rta=\rightarrow
\let\lra=\leftrightarrow
\let\x=\times
\let\dy=\displaystyle

\let\hl=\hfill
\newcommand{\m}{\mbox}
\newcommand{\ol}[1]{\makebox[\textwidth][s]{#1}}

\newcommand{\eqdf}{\stackrel{\mathrm{def}}{=}}
\newcommand{\hf}{\ensuremath{{\scriptstyle\frac{1}{2}}}}
\newcommand{\hfs}{\ensuremath{{\scriptscriptstyle\frac{1}{2}}}}

\newcommand{\be}{\begin{equation}}
\newcommand{\ee}{\end{equation}}
\newcommand{\dd}[3]{\\ \m{}\\ \ol{\m{#1}\hl\m{${\dy #2}$}\hl\m{#3}}\\ \m{}\\}
\newcommand{\re}[2]{\dd{}{#1}{(#2)}}

\newcommand{\ba}{\begin{array}}
\newcommand{\ea}{\end{array}}
\newcommand{\bea}{\begin{eqnarray}}
\newcommand{\eea}{\end{eqnarray}}
\newcommand{\beas}{\begin{eqnarray*}}
\newcommand{\eeas}{\end{eqnarray*}}

\newcommand{\CP}{\mr{CP}}
\newcommand{\qH}{\q{H}}
\newcommand{\qHcp}{\qH_\CP}
\newcommand{\Hfr}{H_{\mr{free}}}
\newcommand{\qHfr}{\q{H}_{\mr{free}}}

\newcommand{\Hnz}{H^{\mr{(NR)}}_{\mr{EM;0}}}
\newcommand{\Hrz}{H^{\mr{(REL)}}_{\mr{EM;0}}}
\newcommand{\Hnh}{H^{\mr{(NR)}}_{\mr{EM;}\hfs}}
\newcommand{\Hrh}{H^{\mr{(REL)}}_{\mr{EM;}\hfs}}
\newcommand{\psid}{\psi^\dg}
\newcommand{\psicp}{\psi_\CP}
\newcommand{\psidcp}{\psi^\dg_\CP}

\newcommand{\vr}{\bv{r}}
\newcommand{\vv}{\bv{v}}
\newcommand{\va}{\bv{a}}
\newcommand{\vp}{\bv{p}}
\newcommand{\vP}{\bv{P}}

\newcommand{\vA}{\bv{A}}

\newcommand{\vz}{\bv{0}}

\newcommand{\qvr}{\q{\vr}}
\newcommand{\qvv}{\q{\vv}}
\newcommand{\qva}{\q{\va}}

\newcommand{\qvp}{\q{\vp}}
\newcommand{\qvP}{\q{\vP}}

\title{The basis of quantum mechanics' compatibility \\
       with relativity---whose impairment gives rise \\
          to the Klein-Gordon and Dirac equations}
\author{Steven Kenneth Kauffmann \\
        American Physical Society Senior Life Member}
\date{43 Bedok Road \\
      {\#}01-11 \\
      Country Park Condominium \\
      Singapore 469564 \\
      Handphone: +65 9370 6583 \\
      \m{} \\
      and \\
      \m{} \\
      Unit 802, Reflection on the Sea \\
      120 Marine Parade \\
      Coolangatta QLD 4225 \\
      Australia \\
      Tel/FAX: +61 7 5536 7235 \\
      Mobile:  +61 4 0567 9058 \\
      \m{} \\
      Email: SKKauffmann@gmail.com}
\begin{document}
\maketitle
\begin{abstract}
Solitary-particle quantum mechanics' inherent compatibility with special rela%
tivity is implicit in Schr\"{o}\-dinger's postulated wave-function rule for the
operator quantization of the particle's canonical three-momentum, taken
together with his famed time-dependent wave-function equation that analogously
treats the operator quantization of its Hamiltonian.  The resulting formally
four-vector equation system assures proper relativistic covariance for any
solitary-particle Hamiltonian operator which, together with its canonical
three-momentum operator, is a Lorentz-covariant four-vector operator.  This, of
course, is \e{always} the case for the quantization of the Hamiltonian of a
properly relativistic \e{classical} theory, so the strong correspondence
principle \e{definitely remains valid} in the relativistic domain.
Klein-Gordon theory \e{impairs} this four-vector equation by \e{iterating} and
contracting it, thereby injecting \e{extraneous} negative-energy solutions that
are \e{not orthogonal} to their positive-energy counterparts of the same
momentum, thus destroying the basis of the quantum probability interpretation.
Klein-Gordon theory, which thus depends on the \e{square} of the Hamiltonian
operator, is as well thereby cut adrift from Heisenberg's equations of motion.
Dirac theory \e{confuses} the space-time symmetry of the \e{four-vector
equation system} with such symmetry for its \e{time component alone}, which it
fatuously \e{imposes}, thereby \e{breaching} the strong correspondence princi%
ple for the free particle and imposing the starkly unphysical \e{momentum-inde%
pendence of velocity}.  Physically sensible alternatives, \e{with} external
electromagnetic fields, to the Klein-Gordon and Dirac equations are derived,
and the simple, elegant symmetry-based approach to antiparticles is pointed out.
\end{abstract}

\subsection*{Introduction: quantum mechanics' inherent compatibility
             with relativity}

The inherent compatibility of solitary-particle quantum mechanics with
special relativity is a straightforward consequence Schr\"{o}dinger's
two basic postulates for the wave function~\ct{Scf,B-D},
i.e., for the quantum state vector in the Schr\"{o}dinger picture in
configuration representation, namely $\la\vr|\psi(t)\ra$.  The first
Schr\"{o}dinger wave-function postulate is his rule for the operator
quantization of the particle's \e{canonical three-momentum},
\re{
    -i\h\del_{\vr}(\la\vr|\psi(t)\ra) = \la\vr|\qvp|\psi(t)\ra,
}{1a}
which is as well, of course, a result of Dirac's postulated
canonical commutation relation~\ct{Dir}.
The second Schr\"{o}dinger wave-function postulate is his famed
\e{time-dependent wave equation}~\ct{Scf,Dir,B-D},
\re{
    i\h\partial(\la\vr|\psi(t)\ra)/\partial t = \la\vr|\qH|\psi(t)\ra,
}{1b}
which formally treats the operator quantization of the particle's
\e{Hamiltonian} in a manner analogous to the way Eq.~(1a) treats
the operator quantization of the particle's canonical three-momentum.
The straightforward theoretical physics \e{implication} of Eqs.~(1a)
and (1b) is simply that the operators $\qvp$ and $\qH$ are the
\e{generators} of the wave function's \e{infinitesimal space and time
translations}, respectively.  Therefore, in anticipation of the
\e{restriction on such generators which special relativity imposes},
these two equations are usefully \e{combined} into the \e{single}
formally \e{four-vector} Schr\"{o}dinger equation for the wave
function,
\re{
    i\h\partial(\la\vr|\psi(t)\ra)/\partial x_{\mu} =
                \la\vr|\q{p^{\mu}}|\psi(t)\ra,
}{1c}
where the \e{contravariant four-vector space-time partial derivative
operator} $\pa/\pa x_{\mu}$ is defined as
$\pa/\pa x_{\mu}\eqdf(c^{-1}\pa/\pa t, -\del_{\vr})$, and the \e{formal}
``contravariant four-vector'' energy-momentum operator $\q{p^{\mu}}$ is
defined as $\q{p^{\mu}}\eqdf(\qH/c, \qvp)$.  Since special relativity 
\e{requires} the contravariant space-time partial derivative four-vector
operator $\pa/\pa x_{\mu}$ to transform between inertial frames in
\e{Lorentz-covariant} fashion, it is apparent from Eq.~(1c) that the
\e{Hamiltonian operator} $\qH$ will be \e{compatible with special
relativity} if it is related to the canonical three-momentum operator
$\qvp$ in such a way that \e{also} makes the energy-momentum operator
$\q{p^{\mu}}$ a contravariant four-vector which transforms between
inertial frames in \e{Lorentz-covariant} fashion.  This property of the
Hamiltonian operator will, of course, be satisfied \e{automatically} if
it is \e{the quantization of the Hamiltonian of a properly relativistic
classical theory}.  \e{Therefore the strong correspondence principle
definitely remains valid in the relativistic domain}.

Now for a completely \e{free} solitary particle of nonzero mass m, the
logic of the Lorentz transformation from its \e{rest frame}, where it
has four-momentum $(mc, \vz)$, to a frame where it has velocity $\vv$
(where $|\vv| < c$) leaves \e{no freedom at all in the choice of its
classical Hamiltonian}.  That Lorentz boost takes this particle's
four-momentum to,
\re{
    (mc(1-|\vv|^2/c^2)^{-\hf},\: m\vv(1-|\vv|^2/c^2)^{-\hf}) =
    (E(\vv)/c,\: \vp(\vv)),
}{2a}
which, together with the \e{identity},
\re{
    mc^2(1-|\vv|^2/c^2)^{-\hf} =
    \sqrt{m^2c^4 + |cm\vv|^2(1 -|\vv|^2/c^2)^{-1}},
}{2b}
implies that,
\re{
E(\vv) = \sqrt{m^2c^4 + |c\vp(\vv)|^2} = \Hfr(\vp(\vv)).
}{2c}
Therefore, for the completely \e{free} relativistic
solitary particle of nonzero mass m, theoretically
systematic, conservative adherence to the \e{strong
correspondence principle} flatly determines the
relativistic Hamiltonian operator to be the
square-root operator,
\re{
\qHfr = \sqrt{m^2c^4 + |c\qvp|^2}.
}{3}
\indent
This conclusion \e{even} extends to the \e{free} spin-$\hf$
particle of nonzero mass: notwithstanding that spin-$\hf$
itself is a nonclassical attribute, the nonrelativistic 
Pauli Hamiltonian operator for such a particle \e{automatic%
ally reduces} to the \e{usual} nonrelativistic \e{purely 
kinetic-energy Hamiltonian} operator in the \e{free-particle}
limit, and one can \e{always} find an inertial frame of reference 
in which a \e{free particle of nonzero mass} is \e{completely
nonrelativistic}, i.e., for the completely \e{free} particle one
can \e{always} find an inertial frame of reference in which the
nonzero-mass relativistic free-particle Hamiltonian operator
$\sqrt{m^2c^4 + |c\qvp|^2}$ \e{arbitrarily} well approximates
$mc^2 + |\qvp|^2/(2m)$, which is the \e{Pauli} free-particle
Hamiltonian operator, offset by the merely \e{constant} rest-mass
energy term $mc^2$.

Since, as we shall shortly see, the Dirac free-particle Hamiltonian
operator \e{is very much at odds with the relativistic free-particle
square-root Hamiltonian operator of} Eq.~(3)~\ct{B-D}, \e{even not%
withstanding the complete compatibility of} Eq.~(3) {with the free-%
particle Pauli theory}, it will be imperative to understand exactly
\e{how and why} the Dirac Hamiltonian operator comes to be \e{in con%
flict with the fundamental requirement of relativistic quantum mechan%
ics}, namely that $\q{p^{\mu}} = (\qH/c, \qvp)$ \e{must transform be%
tween inertial frames as a Lorentz-covariant four-vector}.  First,
however, we turn to analysis of the Klein-Gordon theory, which \e{re%
jects the fundamental quantum mechanical} Eq.~(1c) \e{in that precise
form}, instead \e{substituting in its place} Eq.~(1a) together with a
once-iterated and Lorentz-contracted version of Eq.~(1c).

\subsection*{Klein-Gordon theory's impairment of quantum mechanics}

Very strongly motivated by considerations of perceived calculational
ease, which we briefly discuss in the next section, \e{rather than by
those of quantum mechanics}, Klein, Gordon and Schr\"{o}dinger
\e{rejected} the \e{natural} time-dependent Schr\"{o}dinger equation
of Eq.~(1b) in favor of a  \e{once-iterated} and then {Lorentz-con%
tracted} Lorentz-scalar version of Eq.~(1c)~\ct{B-D, Ka1, Ka2}, which
yields,
\re{
    \pa^2(\la\vr|\psi(t)\ra)/(\pa x^{\mu}\pa x_{\mu}) +
    (\la\vr|\q{p_{\mu}}\,\q{p^{\mu}}|\psi(t)\ra)/\h^2 = 0,
}{4a}
where, of course,
\re{
    \q{p_{\mu}}\,\q{p^{\mu}} = (\qH^2 - |c\qvp|^2)/c^2,
}{4b}
so that \e{in the special case of the free particle}, where
$\qH = \qHfr$, which is given by Eq.~(3), it is readily seen
that the \e{general} Klein-Gordon equation of Eq.~(4a) above
\e{reduces} to,
\re{
    (\pa^2/(\pa x^{\mu}\pa x_{\mu}) + (mc/\h)^2)\la\vr|\psi(t)\ra = 0.
}{5}
To each stationary eigensolution $e^{-i\sqrt{m^2c^4 +
|c\vp |^2}\:t/\h}\la\vr|\vp\ra$ of eigenmomentum $\vp$ of the natural time-%
dependent relativistic \e{free} solitary-particle Schr\"{o}dinger equation,
which is Eq.~(1b) for the case that the Hamiltonian operator $\qH$ is equal
to $\qHfr$, Eq.~(5) \e{adds} an \e{extraneous negative-energy partner
solution} $e^{+i\sqrt{m^2c^4 + |c\vp |^2}\:t/\h}\la\vr|\vp\ra$ of the \e{same}
momentum, whose \e{sole reason for existing} is the \e{entirely gratuitous
iteration} of Eq.~(1c)!  These \e{completely} extraneous \e{negative}
``free solitary-particle'' energies, $-\sqrt{m^2c^4 + |c\vp |^2}$, do
\e{not} correspond to \e{anything} that exists in the \e{classical}
dynamics of a free relativistic solitary particle, and by their negatively
\e{unbounded} character threaten to spawn unstable runaway phenomena
should the \e{free} Klein-Gordon equation be sufficiently perturbed (the
Klein paradox)~\ct{B-D}.
Due to the fact that the Klein-Gordon equation \e{lacks} a corresponding
Hamiltonian---it depends on only the \e{square} of a Hamiltonian, as is seen
from Eq.~(4b)---it turns out, as is easily verified, that the \e{two}
solutions of the \e{same momentum} $\vp$ which have \e{opposite-sign}
energies, i.e., $\pm\sqrt{m^2c^4 + |c\vp |^2}$, \e{fail to be orthogonal to
each other}, which \e{outright violates a key property} of orthodox quantum
mechanics!  \e{Without this property} the probablity interpretation of
quantum mechanics \e{cannot be sustained}, and the Klein-Gordon equation is
unsurprisingly \e{diseased} in that regard, yielding, inter alia,
\e{negative probabilities}~\ct{B-D}.  

Furthermore, Klein-Gordon theory, which depends on the \e{square} of a
Hamiltonian operator (see Eq.~(4b)) rather than on the Hamiltonian
operator \e{itself}, is thereby \e{cut adrift} from the normal quantum
mechanical relationship to the Heisenberg picture, Heisenberg's equations
of motion and the Ehrenfest theorem.  In a nutshell, \e{substitution} of
the \e{gratuitous iteration} of Eq.~(1c) for the Eq.~(1b) \e{part} of
Eq.~(1c), which is \e{exactly} what Klein-Gordon theory \e{does}, grievously
\e{impairs and degrades} the \e{unexceptionable quantum-mechanical nature}
of Eq.~(1c) \e{itself}!  The need to carefully respect \e{all} the compon%
ents of Eq.~(1c) rather than to opportunistically try to ``bend'' aspects of
its time component (i.e., Eq.~(1b)) ``at the edges'' unfortunately \e{did
not register at all} with Dirac ahead of his attempt to ``repair'' the
problems of the Klein-Gordon theory.

\subsection*{Dirac theory's fatuous \e{imposition} of ``space-time coordinate
             symmetry''}

The probability disease of the Klein-Gordon theory prompted Dirac to
sensibly \e{reinstate} Eq.~(1b), the time-dependent Schr\"{o}dinger
equation.  Very unfortunately, being motivated \e{above all else} by
the \e{same non-physics considerations of perceived calculational ease}
as Klein, Gordon and Schr\"{o}dinger~\ct{B-D, Ka2}, Dirac hit upon a
completely misguided but nonetheless \e{plausible-sounding} ``reason'',
ostensibly emanating from the ``relativistic need'' for Eq.~(1b) to
\e{by itself} exhibit ``space-time coordinate symmetry''~\ct{D28, Scw, B-D},
to \e{linearize} the square-root Hamiltonian operator $\qHfr$ of Eq.~(3)
for the nonzero-mass free particle~\ct{B-S, D28, Ka2}, \e{notwithstanding}
the fact that Eqs.~(2) \e{effectively produce the specifically square-root
form of} $\Hfr$ from \e{nothing more} than the nonzero-mass free particle's
Lorentz transformation properties!

Unfortunately \e{not being aware of the specifically four-vector form
of} Eq.~(1c), Dirac focused \e{myopically on only} Eq.~(1b), the
time-dependent Schr\"{o}dinger equation, whose \e{theoretical physics
essence} is, of course, that the Hamiltonian operator is the \e{gen%
erator} of infinitesimal \e{time translations} of the wave function.
This theoretical-physics \e{core essence} of Eq.~(1b) is, in isolation,
\e{supremely indifferent} to the typical space-time coordinate symmetry
that is such an ubiquitous characteristic of special relativity.  It is
\e{only} on contemplating the \e{whole} of the four-vector equation
\e{system} of Eq.~(1c), \e{which Dirac was not aware enough to do}, that
its \e{global} symmetry between space and time coordinates snaps \e{imme%
diately} into focus.  Very unfortunately \e{deprived} of this revelation,
and \e{intently} focused on \e{just} Eq.~(1b), Dirac ``concluded'' that it
must \e{of itelf} be \e{compelled} to manifest the ``missing'' space-time
coordinate symmetry!  If Dirac had even been \e{fully cognizant} of the
\e{core theoretical physics role} of Eq.~(1b), i.e., that of putting the
Hamiltonian operator in the driver's seat of infinitesimal \e{time trans%
lations} and \e{only time translations}, of the wave function, he defin%
itely would have been \e{very much less certain} that \e{forcing} Eq.~(1b)
to manifest space-time coordinate symmetry was \e{at all} a sensible idea!
Such salutary doubts and second thoughts were, however, \e{unfortunately
never to occur to Dirac}.

So because Eq.~(1b) is \e{linear} in the time partial derivative
operator, Dirac shallowly concluded that ``space-time coordinate
symmetry'' \e{compels} it's relativistic version \e{to also be
linear in the space gradient}, which implies that the Hamiltonian
operator on its right-hand side is \e{likewise} linear in the
particle three-momentum.  Had he been thinking \e{at all} about
the \e{physical implications} of this, Dirac would have \e{quickly}
noticed a \e{plethora of red flags} inherent in \e{any such linear%
ity}.  In the nonrelativistic limit of \e{small} particle three-mom%
entum, it is obvious that the Hamiltonian \e{must} become a \e{quad%
ratic} function of the particle three-momentum, which is a \e{mathe%
matical impossibility} for a Hamiltonian \e{linear} in the three-mom%
entum!  Note, however, that the relativistic \e{square-root} Hamilton%
ian $\qHfr$ of Eq.~(3), from whose clutches Dirac, like Klein, Gordon,
Schr\"{o}dinger before him, was very strongly resolved, for non-physics
reasons of calculational ease, to \e{escape}, in fact \e{passes} this
\e{small} three-momentum \e{quadratic-dependence} test with flying col%
ors!  Furthermore, since the square-root Hamiltonian operator's \e{clas%
sical precursor} $\Hfr$ is established beyond \e{any shadow of a doubt
for a nonzero-mass free particle} (e.g., see Eqs.~(2)), Dirac was effect%
ively boxed into the \e{very} fraught position of \e{implicitly disowning
the strong correspondence principle} in order to \e{cling} to his \e{lin%
earized} Hamiltonian!  Furthermore, it is an elementary observation from
the Heisenberg equation of motion that \e{any} Hamiltonian which is
\e{linear} in the particle three-momentum operator produces a particle
\e{velocity} operator (and thus also a particle speed operator) which is
\e{completely independent} of the particle three-momentum operator!  This
obviously \e{cannot make sense} in the free-particle \e{nonrelativistic
limit}, where it is clear that the particle velocity operator \e{must} be
\e{proportional} to the particle momentum operator, i.e., $d\q{\vr}/dt =
\qvp/m$.  That free-particle velocity should be \e{independent} of free-%
particle momentum is in fact totally devoid of \e{physical} sense in
\e{any} regime pertaining to the \e{relativistic} free particle: it is
\e{entirely clear} that relativistic free-particle velocity and momentum
are \e{always directionally aligned} (this \e{even} is true for the ultra-%
relativistic massless photon).

In utter contrast to this miasma of impending \e{physically senseless}
``predictions'' \e{irrespective} of \e{what} the \e{coefficients} of
the three-momentum components in the \e{linearized} Dirac Hamiltonian
are ultimately chosen to be, the Heisenberg equation of motion in
\e{conjunction} with the free-particle relativistic \e{square-root}
$\qHfr$ Hamiltonian operator of Eq.~(3) yields \e{only physically
impeccable} relativistic free-particle results, \e{including}, in
particular, the result for the \e{relativistic velocity operator}!
We shall, of course, be adhering to Dirac's choice of coefficients
for the components of the three-momentum in his linearized Hamiltonian,
but that particular choice \e{doesn't} alleviate \e{any} of the unphys%
ical results just pointed out; in fact, it turns out to \e{sharpen}
them, with the momentum-independent particle \e{speed} coming out to be
a \e{universal c-number} whose value is more than 70\% \e{greater} than
that of light!  As if this were not enough, the Dirac choice of coeffi%
cients (which turns out to involve mutually anticommuting matrices
whose squares are unity) results in a \e{staggering violation of New%
ton's first law of motion}, with the free electron predicted to have a
spontaneous ``Compton acceleration'' whose \e{minimum} magnitude, of
order $m_ec^3/\h$, is around $10^{28} g$, where $g$ is the acceleration
of gravity at the earth surface!  Almost needless to say, the relativis%
tic free-particle \e{square-root} Hamiltonian operator $\qHfr$ of
Eq.~(3), in conjunction with Heisenberg's equation of motion yields, in
\e{flawless} contrast, \e{utterly strict adherence} to Newton's first
law of motion!

In fact, the only obvious ``success'' that the Dirac theory's \e{com%
pletely misdirected} imposition of space-time coordinate symmetry on
Eq.~(1b) can claim is the \e{non-physics} one of \e{calculational ease},
which, as we have pointed out, was its \e{overriding motivation} from
the \e{very start}.  In configuration representation, the \e{physically
sensible} relativistic free-particle \e{square root} Hamiltonian opera%
tor $\qHfr$ of Eq.~(3) is a \e{non-local integral operator}, which cer%
tainly quashes any hoped-for ``separation of variables for partial dif%
ferential equations'' technology for solving the \e{relativistic} ver%
sion of the hydrogen atom.  Contemplation of this fact of physics seems
to have thoroughly rattled Klein, Gordon, Schr\"{o}dinger and, in due
course, Dirac.  Quite \e{desperately} wanting that standard partial dif%
ferential equation technology to \e{still be applicable} in the relati%
vistic domain, they first invented the quantum-mechanically deficient
Klein-Gordon theory and then Dirac's astonishingly unphysical linearized
Hamiltonian.  Schr\"{o}dinger, the \e{inventor} of bound-state perturba%
tion theory, \e{ought to have realized} that his perturbation method is
\e{very well suited} to the patently \e{small} relativistic corrections
to the \e{nonrelativistic model} of the hydrogen atom, and called off
the jointly shared slighting of theoretical-physics best practice which
was tacit in the above-noted quite desperate efforts to \e{preserve the
applicability} of standard partial differential equation solution tech%
nologies in the relativistic domain.  To be sure, the very \e{calcula%
tional ease} that is the key aspect of Dirac's unphysical linearized
Hamiltonian \e{also} makes it a relatively easy target to pick apart in
detail!

We turn now to Dirac's choice of coefficients for his linearized
relativistic mass-$m$ free-particle Hamiltonian operator $\qH_D$
and the consequences of that choice.  Having \e{rejected} the
strong correspondence principle in favor of imposing ``space-time
coordinate symmetry'' on Eq.~(1b) and thus \e{linearity} in the
components of the three-momentum operator on $\qH_D$, Dirac adopt%
ed a \e{severely weakened correspondence principle} for $\qH_D$
in relation to the relativistic free-particle square-root Hamil%
tonian operator $\qHfr$ of Eq.~(3), namely,
\re{
    (\qH_D)^2 = (\qHfr)^2 = m^2c^4 + c^2|\qvp|^2,
}{6a}
which ensures that any solutions of the time-dependent Schr\"{o}%
dinger equation, i.e., Eq.~(1b), with Hamiltonian operator $\qH$
equal to Dirac's relativistic free-particle $\qH_D$ will \e{also}
be solutions of the free-particle Klein-Gordon equation, namely
of Eq.~(5).  Now expressing the linearized $\qH_D$ in terms of
four dimensionless Hermitian matrix coefficients $(\bt, \vec\al)$,
where $\vec\al\eqdf(\al_1, \al_2, \al_3)$, in the form,
\re{
    \qH_D = c\vec\al\dt\qvp + \bt mc^2,
}{6b}
it is seen that the weak correspondence principle of Eq.~(6a)
implies that $\bt$ and all the components of $\vec\al$ mutually
\e{anticommute} and have squares equal to unity.  Their anti%
commutation relations imply that they are traceless~\ct{B-D},
and therefore $\qH_D$ is traceless as well, and thus has a
\e{negative} energy eigenvalue to match every positive one.
So the eigenenergies of Dirac's linearized $\q{H}_D$ turn out
to \e{include} all the extraneous \e{negative} energies of the
Klein-Gordon equation.  While the negative-energy eigenstates
of $\qH_D$ \e{are} properly \e{orthogonal} to their positive-%
energy counterparts, the \e{other} inherent issues which the
presence of these negative-energy solutions raise in the context
of a free solitary particle, such as total lack of classical cor%
respondence and the Klein paradox remain unresolved~\ct{Ka1}.
Upon applying the Heisenberg equations of motion to the
free-particle Dirac Hamiltonian operator $\qH_D$, as given by
Eq.~(6b), we obtain the velocity operator,
\re{
    \qvv = d\qvr/dt = c\vec\al,
}{7a}
which we see, as pointed out above, is \e{completely independent}
of the three-momentum operator $\qvp$.  Since the three Hermitian
matrix components of $\vec\al$ all have squares equal to unity, we
obtain for the free Dirac particle speed operator the \e{universal}
c-number value,
\re{
    |\qvv| = c\sqrt{3},
}{7b}
which \e{exceeds} $c$, the speed of light, by 73\%!  Also, notwith%
standing its nonzero mass $m$, the free Dirac particle \e{has no
inertial rest frame}!  These results \e{show conclusively} that the
\e{imposition} on Eq.~(1b) of ``space-time coordinate symmetry''
with its resulting \e{linearized} Hamiltonian operator, that is
\e{crystallized} as the free-particle Dirac Hamiltonian operator
$\qH_D$ of Eq.~(6b) upon imposition of the severely weakened corres%
pondence principle of Eq.~(6a), \e{violates} special relativity!

To better grasp this point, let us note that if we use the
relativistic square-root Hamiltonian operator $\qHfr$ \e{instead}
of $\qH_D$ in the Heisenberg equations of motion, we obtain,
\[\qvv = c^2\qvp/(m^2c^4 + |c\qvp|^2)^\hf,\]
which is a momentum-dependent velocity operator whose magnitude is
always strictly smaller than the speed of light $c$, and which will
have the value zero for an eigenstate of $\qvp$ which has three-mom%
entum eigenvalue equal to $\vz$.  So the particle rest frame certainly
exists.  We are actually in a position to now \e{repeat} the Eqs.~(2)
Lorentz boost of a nonzero-mass free particle's four-momentum from its
rest frame to an inertial frame in which it has nonvanishing velocity,
but this time \e{using for that velocity the operator value} $\qvv$
\e{that arises from $\qHfr$}, which \e{operator} $\qvv$ is displayed
just above, \e{instead of using for that velocity the c-number value}
$\vv$ of Eqs.~(2).  If we \e{simply follow} the steps of Eqs.~(2),
albeit using this particular \e{operator} $\qvv$, we end up with the
\e{operator four-momentum} $((m^2c^2 + |\qvp|^2)^\hf , \qvp)$, which
\e{accords perfectly} with the \e{three-momentum operator} $\qvp$ and
the \e{Hamiltonian operator} $(m^2c^4 + |c\qvp|^2)^\hf$.

Now let us try to \e{emulate} the above successful ``Lorentz boost
of particle four-momentum out of the particle rest frame using the
\e{operator} $\qvv$ that arises from $\qHfr$'' exercise by \e{in%
stead} using the simple operator $\qvv$ that arises from the
\e{Dirac} $\qH_D$, which of course is $\qvv = c\vec\al$.  For
$\qvp = \vz$, we have that $\qH_D = \bt mc^2$.  If we now try to
follow the steps of Eqs.~(2) from this starting point by using the
Dirac $\qvv = c\vec\al$, we end up with the quite senseless operator
four-momentum $(\bt mc(-i/\sqrt{2}\:), \bt\vec\al mc(-i/\sqrt{2}\:))$
that bears little resemblance to the \e{desired} operator four-momen%
tum result $(\vec\al\dt\qvp + \bt mc, \qvp)$.  In other words, for the
Dirac theory Hamiltonian operator $\qH_D$, the attempt to show Lorentz
covariance has ended in disaster.  The Dirac theory contravenes special
relativity just as the Klein-Gordon theory contravenes quantum mechanics.

We have pointed out that for the free-particle nonrelativistic Pauli
theory the Hamiltonian operator has only the kinetic energy term
$|\qvp|^2/(2m)$, which implies that \e{orbital} angular momentum
$\qvr\x\qvp$ is \e{exactly} conserved.  In contrast, the free-particle
Dirac theory has a very strong spin-orbit coupling which almost
certainly is not physically sensible.  From Eq.~(7a) it is clear that
in the free-particle Dirac theory,
\re{
    d(\qvr\x\qvp)/dt = c\vec\al\x\qvp,
}{8a}
and therefore the magnitude of the spin-orbit \e{torque} is,
\re{
    |d(\qvr\x\qvp)/dt| = c|\qvp|\sqrt{2}\:,
}{8b}
Now the particle's kinetic energy is $((m^2c^4 + |c\qvp|^2)^\hf - mc^2)$.
If we take the \e{dimensionless ratio} of the particle's spin-orbit
torque magnitude to its kinetic energy, we obtain
$((1 + (mc/|\qvp|)^2)^\hf + (mc/|\qvp|))\sqrt{2}\:$, which increases
monotonically \e{without bound} as $|\qvp|$ \e{decreases}.  This
is, of course, \e{not consistent} with the free-particle Pauli
theory, where this ratio always \e{vanishes identically}.  So the
Dirac theory \e{does not} reduce to the Pauli theory merely by
going to small values of momentum.  This was already clear, of
course, from the fact that the Dirac particle's \e{speed} always
has the value $c\sqrt{3}$ \e{irrespective} of its momentum, which
\e{doesn't accord} with the free-particle Pauli theory at all.

We now present the details of the free Dirac particle's stagger%
ing violation of Newton's first law of motion.  Using Heisenberg's
equations of motion, the Dirac free-particle Hamiltonian $\qH_D$
operator of Eq.~(6b), and the simple Dirac free particle velocity
operator $\qvv = c\vec\al$, we obtain the Dirac free-particle
spontaneous acceleration operator,
\re{
    \qva = d\qvv/dt = (2mc^3/\h)(i\bt\vec\al + (\qvp\x\vec\sg)/(mc)),
}{9a}
where,
\re{
    \vec\sg\eqdf(-i/2)(\vec\al\x\vec\al),
}{9b}
which means,
\re{
    \sg_i\eqdf(-i/2)\ep_{ijk}\al_j\al_k.
}{9c}
Since $|i\bt\vec\al|^2 = 3$, $|\qvp\x\vec\sg|^2 = 2|\qvp|^2$, and
$(i\bt\vec\al)\dt(\qvp\x\vec\sg) = -(\qvp\x\vec\sg)\dt(i\bt\vec\al)$,
we obtain for the magnitude of the spontaneous acceleration,
\re{
    |\qva| =  (2\sqrt{3}\:mc^3/\h)(1 + (2/3)(|\qvp|/(mc))^2)^\hf,
}{9d}
whose minimum value, $(2\sqrt{3}\:mc^3/\h)$, is, for the case of the
electron, well in excess of $10^{28} g$, where $g$ is the acceler%
ation of gravity at the earth's surface.  This dumbfounding spontan%
eous acceleration certainly \e{drives home the point} that the
``free'' Dirac particle \e{has no inertial rest frame}.  Note that
matters \e{don't even improve} as the Dirac particle is made more
massive; the minimum spontaneous acceleration is \e{proportional}
to the particle mass!  The systematics of this unphysical absurdity
show \e{just how profoundly} the Dirac theory \e{bungles special
relativity}. As previously mentioned, the relativistic free-particle
\e{square-root} Hamiltonian operator $\qHfr$ \e{strictly adheres} to
Newton's first law of motion, yielding $\qva = \vz$.

Finally, it is a \e{built-in property} of special relativity that
the character of the physics it describes becomes Galilean/Newton%
ian if the speed of light $c$ is taken to be \e{asymptotically
large}.  For example, if we subtract from the relativistic free-%
particle square-root Hamiltonian operator $\qHfr$ of Eq.~(3) its
value at $\qvp = \vz$, which is $mc^2$, the limit of that differ%
ence as $c\rta\infty$ is, of course, the \e{nonrelativistic free-%
particle kinetic energy operator} $|\qvp|^2/(2m)$.  But if we carry
out \e{exactly same steps} with the Dirac Hamiltonian operator $\qH_D$,
we face \e{formal divergence} as $c\rta\infty$!  Likewise, if we take
the particular velocity operator $\qvv$ discussed in the paragraph
\e{which follows} the one in which Eq.~(7b) occurs, which is obtained
from the Heisenberg equation of motion \e{in conjunction with}
$\qHfr$, and is explicitly given by,
\[\qvv = c^2\qvp/(m^2c^4 + |c\qvp|^2)^\hf,\]
it is clear that as $c\rta\infty$, $\qvv\rta\qvp/m$, i.e., \e{in this
limit $\qvv$ becomes the nonrelativistic free-particle velocity opera%
tor}.  On the other hand, for the Dirac free-particle velocity opera%
tor $\qvv$ of Eq.~(7a) we \e{again} face \e{formal divergence} as
$c\rta\infty$!  The clear lesson from these instances is that a
\e{linearized} Hamiltonian operator such as the Dirac $\qH_D$ is \e{in%
herently incapable of incorporating the physics of special relativity}.
It is obvious the \e{mathematical presence of a square root} is \e{es%
sential} to correctly capturing the \e{the properties of relativistic
physics}.  \e{Even} in the case of the Klein-Gordon equation, which
``squares out'' the mathematical presence of the square root, it is
\e{impossible} to show that the ``physics'' it describes becomes prop%
erly nonrelativistic as $c\rta\infty$.  Looking in detail at the solu%
tion space of the Klein-Gordon equation, we readily see that this is
due the \e{gratuitous presence} of of the \e{extraneous} negative-ener%
gy solutions, which are, of course, \e{unphysical detritus}.

\subsection*{Correct relativistic quantum mechanics with an external
             electromagnetic field}

It is clear that, for the nonzero-mass solitary relativistic free particle,
the Klein-Gordon theory, which disrupts its quantum mechanics, and the
Dirac theory, which dumbfoundingly abolishes its rest frame, \e{must both
be abandoned in favor of the straightforward square-root Hamiltonian opera%
tor} $\qHfr$ of Eq.~(3).  For the free photon, which is massless, the time-%
dependent Schr\"{o}dinger equation (i.e, Eq.~(1b)) \e{in conjunction with
the zero-mass case of} $\qHfr$ turns out to be \e{already implicit in the
source-free case of Maxwell's equations}~\ct{Ka2}.

We shall now develop the \e{extension} of $\qHfr$ to the case of a par%
ticle of charge $e$ and nonzero mass $m$ when an external electromagnet%
ic potential $A^{\mu}(\vr, t)$ is present, first for a spin-0 particle
and then for a spin-$\hf$ particle.  For additional background and detail
concerning the derivation of those two Hamiltonians see reference~\ct{Ka1}.
The guiding concept is that accurate understanding of the physics
experienced by a nonzero-mass solitary particle in an inertial frame
where it is instantaneously traveling arbitrarily slowly translates, via
a continuous sequence of successive Lorentz transformations, into the
accurate understanding of that physics in an \e{arbitrary inertial frame}.
Therefore a tested and trusted \e{nonrelativistic} theory of a nonzero-%
mass solitary particle's behavior ought to always be reasonably straight%
forwardly upgradable to a correct \e{relativistic} one that explicitly
reduces to the underlying nonrelativistic one in any inertial frame in
which that solitary particle happens to be instantaneously moving at non%
relativistic speed.  The approach to attempting to carry out such a pro%
gram which is followed here is to try to associate \e{individual terms}
of the solitary particle's \e{nonrelativistic Hamiltonian} with fully Lor%
entz-covariant energy-momentum four-vectors whose \e{time components re%
duce to those individual nonrelativistic Hamiltonian terms} in inertial
frames where that particle is traveling at \e{arbitrarily slow speed}.  The
\e{individual} Lorentz-covariant energy-momentum four-vectors so determined
are then, of course, \e{summed} to produce the solitary particle's Lorentz-%
covariant \e{relativistic total energy-momentum four-vector}.  The resulting
solitary-particle \e{relativistic total three-momentum} is obviously identi%
fied as the \e{generator of the solitary particle's space translations} and
thus as the solitary particle's \e{relativistic canonical three-momentum}.
Of course the solitary particle's \e{relativistic total energy}, when ex%
pressed as \e{function of its relativistic canonical three-momentum, the
time, and that particle's three space coordinates} comprises that particle's
\e{relativistic Hamiltonian}.  Initially, of course, the \e{individual terms}
contributing to the solitary particle's relativistic total energy-momentum
four-vector will be couched in the language of its three space coordinates,
the time, and that particle's relativistic \e{kinetic} three-momentum.
Upon \e{identification} of the particle's relativistic \e{canonical} (i.e.,
\e{total}) three-momentum, it is then \e{necessary to solve} for its relati%
vistic \e{kinetic} three-momentum as a \e{function} of that relativistic
canonical three-momentum \e{in order to be able to reexpress its total rela%
tivistic energy as its Hamiltonian}.  Regrettably, a very conceivable ``fly
in the ointment'' is \e{that there is no guarantee} that the particle's
relativistic \e{kinetic} three-momentum can be obtained as a \e{function of
its relativistic canonical three-momentum in closed form}.  Thus the soli%
tary particle's relativistic Hamiltonian \e{itself} may conceivably \e{only
be available as a sequence of approximations}.  Such a state of affairs is
obviously \e{not} what one would desire, but \e{unlike} Klein, Gordon,
Schr\"{o}dinger and Dirac, the \e{physically-motivated}, conservative
theorist is, like Einstein, obliged to coexist with whatever undesirable
\e{calculational} consequences that physically-based, conservative theory
happens to carry with it.  The art of the \e{appropriate approximation},
one carefully attuned to a particular problem at hand, is surely yet
another vital skill the theoretical physicist is called upon to hone.

Let us now apply the the above program to a spin-0 solitary particle of
mass $m$ and charge $e$ in the presence of an external electromagnetic
potential $A^\mu(\vr, t)$.  It will be recalled that all \e{magnetic}
effects of such a potential on the particle's motion \e{vanish entirely}
in the particle's rest frame, and are, more generally, of order $O(1/c)$,
whereas in nonrelativistic physics the speed of light $c$ is regarded as
an asymptotically large parameter.  Thus the \e{strictly nonrelativistic}
Hamiltonian operator for this particle involves \e{only} the electromag%
netic potential's time component $A^0(\vr, t)$,
\re{
    \q\Hnz = |\qvp|^2/(2m) + eA^0(\qvr, t).
}{10a}
Because of the technical issue regarding the choice of ordering of
noncommuting operators (whose resolution we allude to below), it will
be convenient to develop the relativistic energy-momentum four-vector as
a function of \e{classical} $(\vr, \vp)$ phase space \e{rather} than as
a function of the \e{already quantized} $(\qvr, \qvp)$ phase space of
Eq.~(10a).  The solitary particle's nonrelativistic kinetic energy
$|\vp|^2/(2m)$, plus its rest mass energy $mc^2$, is well-known to cor%
respond to $c$ times its Lorentz-covariant \e{free-particle kinetic ener%
gy-momentum four-vector} $p^\mu$,
\[p^\mu\eqdf((m^2c^2 + |\vp|^2)^\hf, \vp),\]
where, of course, $\vp$ is the particle's relativistic \e{kinetic}
three-momentum, which was \e{distinguished} in the above discussion
from its relativistic \e{total} (i.e., \e{canonical}) three-momentum.
It is apparent that in the nonrelativistic limit $|\vp|\ll mc$, the time
component times $c$ of $p^\mu$ does indeed, as just mentioned, behave as,
\[ cp^0 \approx mc^2 + |\vp|^2/(2m).\]
The \e{potential energy} term $eA^0(\vr, t)$ of $\Hnz$, divided by $c$,
is obviously the time component of the Lorentz-covariant energy-momentum
four-vector $eA^\mu(\vr, t)/c$.  Therefore adding $eA^\mu/c$ to $p^\mu$
produces a fully Lorentz-covariant \e{total} energy-momentum four-vector
whose time component times $c$ reduces, in any inertial frame in which the
nonzero-mass charged spin-0 solitary particle instantaneously has arbitrar%
ily small speed, to that particle's nonrelativistic classical Hamiltonian
$\Hnz$ (which corresponds to the quantized Hamiltonian operator $\q\Hnz$
of Eq.~(10a)) plus that particle's rest mass energy $mc^2$.  We therefore
regard,
\re{
    P^\mu\eqdf p^\mu + eA^\mu/c,
}{10b}
as that solitary particle's \e{fully relativistic} energy-momentum four-%
vector.  Thus the particle's \e{relativistic total three-momentum} is,
\re{
    \vP = \vp + e\vA(\vr, t)/c,
}{10c}
and its \e{relativistic total energy} is,
\re{
    E(\vr, \vp, t) = cP^0 = (m^2c^4 + |c\vp|^2)^\hf  + eA^0(\vr, t).
}{10d}
Here we are in the fortunate position of being able to \e{solve} Eq.~(10c)
for the particle's relativistic \e{kinetic} three-momentum $\vp$ as a
\e{function} of its relativistic \e{total} (i.e., \e{canonical}) three-mom%
entum $\vP$ in \e{closed form},
\re{
    \vp(\vP) = \vP - e\vA(\vr, t)/c,
}{10e}
which result for $\vp(\vP)$, i.e., $\vp$ \e{as a function of} $\vP$,
we must now \e{substitute} into Eq.~(10d) for the relativistic
\e{total energy} in order to \e{reexpress} that total energy as the
\e{relativistic Hamiltonian} $\Hrz(\vr, \vP, t)$, i.e.,
\[\Hrz(\vr, \vP, t)\eqdf E(\vr, \vp(\vP), t).\]
Therefore Eqs.~(10d) and (10e) yield the following \e{fully relati%
vistic classical Hamiltonian} $\Hrz(\vr, \vP, t)$, \e{which corres%
ponds to our original nonrelativistic Hamiltonian operator} $\q\Hnz$
\e{of} Eq.~(10a),
\re{
 \Hrz(\vr, \vP, t) = (m^2c^4 + |c\vP - e\vA(\vr, t)|^2)^\hf  + eA^0(\vr, t).
}{10f}
\indent
Because of the \e{presence of the square root} in Eq.~(10f) for
$\Hrz(\vr, \vP, t)$, there could conceivably be an issue regarding
the \e{ordering} of the mutually noncommuting operators $\qvr$ and
$\qvP$ when one attempts \e{quantize} this classical Hamiltonian
$\Hrz(\vr, \vP, t)$ to become the \e{Hamiltonian operator} $\q\Hrz$.
Use of the \e{Hamiltonian phase-space path integral}~\ct{Ka3} with
$\Hrz(\vr, \vP, t)$ in its \e{classical form as given by} Eq.~(10f)
provides one \e{definitive solution} to any such operator-ordering
issue.  Another \e{completely equivalent} solution to this issue
lies with a \e{natural slight strengthening} of Dirac's canonical
commutation rule such that \e{it remains self-consistent}~\ct{Ka4}.
From \e{both} of these approaches the resulting \e{unambiguous opera%
tor-ordering rule} turns out to be the one of Born and Jordan~\ct{B-J}.

It is well worth noting that the relativistic classical Hamiltonian
$\Hrz(\vr, \vP, t)$ of Eq.~(10f) for the solitary spin-0 charged
particle, when inserted into Hamilton's classical equations of motion,
yields, upon taking proper account of Eq.~(10c), \e{the fully relativis%
tic version of the Lorentz-force law}.  In other words, the Hamiltonian
$\Hrz(\vr, \vP, t)$ embodies \e{nothing more or less} than the well-known
classical relativistic physics of the charged particle developed by
H.\ A.\ Lorentz~\ct{L-L}.  It is \e{truly a lamentable pity} that Klein,
Gordon and Schr\"{o}dinger went off on their \e{tangent which plays such
havoc with quantum mechanics} instead of simply \e{going forward in work%
manlike fashion with the quantization of Lorentz' utterly transparent rel%
ativistic legacy}, namely the $\Hrz(\vr, \vP, t)$ of Eq.~(10f) above.
Of course, upon taking the particle charge $e$ to zero,
$\Hrz(\vr, \vP, t)$ simply becomes the $\Hfr$ of Eq.~(2c), as it indeed
must.

We turn now to the spin-$\hf$ charged particle, whose nonrelativistic
Hamiltonian $\Hnh$ is the \e{same} as $\Hnz$ \e{except for an additional
interaction energy between the particle's intrinsic magnetic moment
and the external magnetic field}, i.e., its Pauli energy.  Notwith%
standing that this Pauli magnetic dipole energy is customarily
\e{formally written} as including a factor of $(1/c)$, \e{it must
nonetheless be kept in the nonrelativistic limit} because it \e{fails
to vanish} in the spin-$\hf$ particle's \e{rest frame},
\re{
    \Hnh = |\vp|^2/(2m) + 
           (ge/(mc))(\h /2)\vec\sg\dt(\del_\vr\x\vA(\vr ,t))
           + eA^0(\vr, t).
}{11a}
The Pauli magnetic dipole energy contribution to $\Hnh$ is the \e{only
part} of this Hamiltonian which is \e{not a multiple of the two-by-two
identity matrix} in the spin-$\hf$ two-by-two-matrix degrees of freedom.
Now the relativistic energy-momentum four-vectors we shall be developing
will of course \e{themselves} be two-by-two matrices, but this should not
present difficulties so long as their four components \e{all mutually
commute}.  To \e{ensure} that this is the case, we shall ``quarantine''
the non-identity Pauli energy matrix into a \e{Lorentz scalar}, which we
can furthermore render \e{dimensionless} by dividing by $mc^2$.  If we
now \e{multiply} this \e{dimensionless Lorentz scalar} by the particle's
\e{kinetic} energy-momentum four-vector $p^\mu = ((m^2c^2 + |\vp|^2)^\hf,
\vp)$, we will indeed have the desired energy-momentum contribution whose
\e{time component times} $c$ reduces to the Pauli energy matrix in the
particle rest frame.

There remains, of course, the challenging problem of turning the compli%
cated Pauli energy term into a Lorentz scalar.  In relativistic tensor
language, the magnetic field axial vector $(\del_\vr\x\vA(\vr, t))$ that
appears in the Pauli term comprises a certain \e{three-dimensional part}
of the four-dimensional relativistic second-rank antisymmetric electromag%
netic \e{field tensor} $F^{\mu\nu}(\vr, t) = \pa^\mu A^\nu(\vr, t) -
\pa^\nu A^\mu(\vr, t)$.  If we can manage to reexpress the three-dimension%
al axial spin-$\hf$ vector $(\h /2)\vec\sg$ as a ``matching'' three-dimen%
sional part of a four-dimensional relativistic second-rank antisymmetric
tensor $s^{\mu\nu}$ as well, hopefully the Pauli energy will end up being
proportional to to their \e{Lorentz-scalar contraction}
$s^{\mu\nu}F_{\mu\nu}(\vr, t)$.  As a first step, we define the natural
\e{three-dimensional} second-rank antisymmetric spin-$\hf$ tensor $S^{ij}$
in terms of the spin-$\hf$ axial vector,
\[S^{ij}\eqdf (\h /2)\ep^{ijk}\sg^k,\]
and take note that the complicated factor in the Pauli energy term neatly
reduces to a contraction of $S^{ij}$ with the well-known \e{magnetic-field}
three-dimensional part $F^{ij}(\vr, t)$ of $F^{\mu\nu}(\vr, t)$,
\[(\h /2)\vec\sg\dt(\del_\vr\x\vA(\vr ,t)) = (1/2)S^{ij}F^{ij}(\vr, t),\]
which allows us to reexpress the nonrelativistic Eq.~(11a) in the
\e{relativistically more suggestive form},
\re{
    mc^2 + \Hnh = mc^2[1  + |\vp|^2/(2m^2c^2) + 
                  (g/2)(e/(m^2c^3))S^{ij}F^{ij}(\vr, t)]
                  + eA^0(\vr, t).
}{11b}
\indent
It is now time to work out the nonzero-mass spin-$\hf$ particle's fully
covariant \e{four-dimensional} antisymmetric spin tensor $s^{\mu\nu}$.
In the \e{particle rest frame}, namely in the special inertial frame
where the \e{particle kinetic three-momentum} $\vp$ \e{vanishes}, the nine
space-space components of $s^{\mu\nu}$ must clearly be the nine components
of $S^{ij}$, and its remaining seven components must be \e{filled out with
zeros}, i.e.,
\[s^{\mu\nu}(\vp = \vz)\eqdf  \dl^\mu_i\dl^\nu_jS^{ij},\]
which \e{ensures} that, in the particle rest frame,
\[s^{\mu\nu}(\vp = \vz)F_{\mu\nu}(\vr, t) = S^{ij}F^{ij}(\vr, t).\]
Once a tensor is \e{fully determined} in \e{one} inertial frame, it is
fully determined in \e{all} inertial frames by application of the appro%
priate Lorentz transformation to its indices.  To get from the particle
rest frame to the inertial frame where the particle has kinetic three-%
momentum $\vp$ simply requires the appropriate Lorentz-boost four-dimen%
sional matrix $\Lambda^\mu_\al(\vv(\vp)/c)$ that is characterised by the
corresponding dimensionless scaled relativistic particle velocity,
\[\vv(\vp)/c = (\vp/(mc))/(1 + |\vp/(mc)|^2)^\hf,\]
and its accompanying dimensionless time-dilation factor,
\[\gm(\vp) = (1 + |\vp/(mc)|^2)^\hf,\]
so that, in general,
\[s^{\mu\nu}(\vp) = \Lambda^\mu_i(\vv(\vp)/c)
                    \Lambda^\nu_j(\vv(\vp)/c)S^{ij},\]
which, of course, \e{ensures} that $s^{\mu\nu}(\vp)F_{\mu\nu}(\vr, t)$
is a Lorentz scalar that \e{Lorentz-invariantly carries the particle's
rest-frame value of} $S^{ij}F^{ij}(\vr, t)$.

With that, we are in the position to be able to write down the Lorentz-%
covariant \e{total} energy-momentum four-vector $P^\mu$ for the spin-$\hf$
particle that corresponds to its nonrelativistic Eq.~(11b) in the \e{same
way} that the total energy-momentum four-vector of Eq.~(10b) for the spin-0
particle corresponds to \e{its} nonrelativistic Eq.~(10a),
\re{
    P^\mu\eqdf p^\mu[1 + (g/2)(e/(m^2c^3))s^{\al\bt}(\vp)F_{\al\bt}(\vr, t)]
               + eA^\mu(\vr, t)/c.
}{11c}
From $P^\mu$ we obtain the particle's relativistic total energy,
\re{
E(\vr ,\vp ,t) = cP^0 = (m^2c^4 +|c\vp |^2)^\hf
[1 + (g/2)(e/(m^2c^3))s^{\mu\nu}(\vp )F_{\mu\nu}(\vr ,t)]
+ eA^0(\vr ,t),
}{11d}
and also its relativistic total three-momentum,
\re{
\vP = \vp [1 + (g/2)(e/(m^2c^3))s^{\mu\nu}(\vp )F_{\mu\nu}(\vr ,t)]
+ e\vA (\vr ,t)/c.
}{11e}
It is obvious from Eq.~(11e) that we \e{cannot solve} for $\vp (\vP )$
in \e{closed form}, but we \e{can} write $\vp (\vP )$ in
``iteration-ready'' form as,
\re{
\vp (\vP ) = (\vP - e\vA (\vr ,t)/c)
[1 + (g/2)(e/(m^2c^3))s^{\mu\nu}(\vp (\vP ))F_{\mu\nu}(\vr ,t)]^{-1},
}{11f}
and, of course, from $E(\vr, \vp(\vP), t)$, we also obtain the
schematic form of the particle's relativistic Hamiltonian,
\re{
\Hrh(\vr ,\vP ,t) =
(m^2c^4 +|c\vp (\vP )|^2)^\hf
[1 + (g/2)(e/(m^2c^3))s^{\mu\nu}(\vp (\vP ))F_{\mu\nu}(\vr ,t)]
+ eA^0(\vr ,t).
}{11g}
If we take the limit $g\rta 0$ in Eqs.~(11f) and (11g), then
$\Hrh(\vr ,\vP ,t)\rta\Hrz(\vr ,\vP ,t)$, as is easily checked
from Eq.~(10f).  Of course it is nothing more than the \e{most
basic common sense} that fully relativistic spin-$\hf$ theory
simply reduces to fully relativistic spin-0 theory when the
spin coupling of the single particle to the external field is
switched off, but analogous cross-checking between the Dirac
and Klein-Gordon theories is never so much as discussed!  It
is certainly possible to add a term to the Dirac Hamiltonian
that \e{cancels out} it's \e{supposed} $g = 2$ spin coupling to
the magnetic field, but the result of doing this bears \e{very
little resemblance} to the Klein-Gordon equation in the presence
of the external electromagnetic potential!  Elementary consis%
tency checks are obviously \e{not} the strong suit of those two
``theories''!

It is unfortunate that Eq.~(11f) for $\vp (\vP )$ is not amenable
to closed-form solution, but if we assume that the spin coupling term,
$(g/2)(e/(m^2c^3))s^{\mu\nu}(\vp (\vP ))F_{\mu\nu}(\vr ,t)$, which is
a dimensionless Hermitian two-by-two matrix, effectively has the
magnitudes of both of its eigenvalues much smaller than unity (which
should be a very safe assumption for atomic physics), then we can
approximate $\vp (\vP )$ via successive iterations of Eq.~(11f), which
produces the approximation $(\vP - e\vA (\vr ,t)/c)$ for $\vp (\vP )$
through zeroth order in the spin coupling and,
\[\vp (\vP )\approx  (\vP - e\vA (\vr ,t)/c)
[1 + (g/2)(e/(m^2c^3))s^{\mu\nu}(\vP - e\vA (\vr ,t)/c)
F_{\mu\nu}(\vr ,t)]^{-1},\]
through first order in the spin coupling.  We wish to
interject at this point that since $s^{\mu\nu}(\vp (\vP ))$
is an antisymmetric tensor, the tensor contraction
$s^{\mu\nu}(\vp (\vP ))F_{\mu\nu}(\vr ,t)$ is equal to
$2s^{\mu\nu}(\vp (\vP ))\partial_{\mu}A_{\nu}(\vr ,t)$,
which is often a more transparent form. Now if we simply
use the approximation $(\vP - e\vA (\vr ,t)/c)$
through zeroth order in the spin coupling for $\vp (\vP )$,
we obtain the following approximation to $\Hrh$,
\re{
\Hrh(\vr ,\vP ,t)\approx
(m^2c^4 +|c\vP - e\vA (\vr ,t)|^2)^\hf
[1 + (ge/(m^2c^3))s^{\mu\nu}(\vP - e\vA (\vr ,t)/c)
\partial_{\mu}A_{\nu}(\vr ,t)] + eA^0(\vr ,t).
}{11h}
\m{}
\vspace{-1.5pc}

\subsection*{Antiparticle partners from field theoretic symmetry}

If one were to write down a purely electromagnetic (and thus
parity-conserving) quantum field theory that treats \e{only}
positive-energy electrons and photons, which would be possible
via second quantization of $\q\Hrh$ and quantization of the
electromagnetic field, it is clear that one would then have a
quantum field theory which exhibits no invariance whatsoever
under charge conjugation.  Simply imposing charge conjugation
invariance on such a theory \e{forces positrons to exist}.
This is a \e{very familiar picture indeed}: the imposition of
symmetries on quantum field theories \e{not at all infrequent%
ly forces families of particles to exist}.  Note that the
positrons that would be forced into existence by the imposi%
tion of charge conjugation invariance would \e{also be purely
positive-energy particles}, at least when free.  ``Reinterpret%
ation'' of unbounded-below, free-particle negative-energy
spectra plays \e{no role whatsoever} in the existence of these
positrons: their existence is driven theoretically \e{entirely
by the enforcement of the symmetry}!

The imposition of charge conjugation invariance---or, more
generally, CP invariance---on solitary-particle quantum
mechanics that has been taken to the quantum field theory
level does not \e{of itself}, however, \e{ensure} that
external fields can mediate the basic physical processes of
particle-antiparticle pair production and annihilation~\ct{Ha},
processes which \e{cannot}, of course, \e{be accommodated in
the context of a strictly solitary-particle idealization}.  To
accommodate these processes, the imposition of CP invariance at
the field theory level \e{needs to be accompanied} by the \e{ad%
ditional} imposition of invariance under the interchange of par%
ticle absorption/emission with, respectively, antiparticle emis%
sion/absorption, which we term ``CP-\e{equivalence}'' symmetry.
To give one crude, schematic sketch of an imposition of the dual
symmetries of CP invariance and ``CP-equivalence'', suppose we
have in hand the quantum mechanics Hamiltonian operator $\qH$ for
a solitary relativistic charged particle interacting with external
fields.  The straightforward second quantization of $\qH$ produces
the quantum-field-theoretic schematic quantized Hamiltonian
density $\psid\qH\psi$, which has the particle emission field
$\psid$ and the particle absorption field $\psi$, but \e{no anti%
particle fields}, and therefore is completely CP \e{noninvariant}.
This is readily remedied by introducing the antiparticle emission
field $\psidcp$ and the antiparticle absorption field $\psicp$,
which correspond to the CP-transformed version of $\qH$ that we
denote as $\qHcp$ (a prototypical difference between $\qH$ and
$\qHcp$ is, of course, in the sign of the solitary particle's
charge).  Using these entities, one simple way to impose CP invari%
ance symmetry is to replace the Hamiltonian density $\psid\qH\psi$
by $(\psid\qH\psi + \psidcp\qHcp\psicp)$, which can permit external
fields to mediate both particle and antiparticle scattering, but not
yet particle-antiparticle pair production or annihilation.  To in%
clude the latter physical phenomena we shall \e{additionally} impose
the ``CP-equivalence'' symmetry which was pointed out above by
\e{also} requiring the Hamiltonian density to be invariant under the
particle field interchanges $\psi\lra\psidcp$ and $\psid\lra\psicp$.
It is apparent that one straightforward upgrade of the solitary-par%
ticle Hamiltonian density $\psid\qH\psi$ which incorporates the
\e{dual symmetries} of CP invariance and ``CP-equivalence'' that we
are stipulating here is given by,
\[\psid\qH\psi\rta(\psi + \psidcp)^\dg\qH(\psi + \psidcp) +
                  (\psicp + \psid)^\dg\qHcp(\psicp + \psid),\]
which clearly can permit external fields to mediate pair production
and annihilation, as well as particle and antiparticle scattering.

The role of symmetries in accounting for and compelling the
existence of energy degeneracies has been a robust and \e{high%
ly fruitful} theme of quantum physics since its \e{very earli%
est days}; the names of such pioneers as Wigner and Weyl come
readily to mind.  The existence of antiparticles is a \e{clas%
sic example of energy degeneracy} in the context of quantum
field theory, \e{and there is absolutely no reason whatsoever
that it should not take its rightful place in the theoretical
physics pantheon of symmetry-driven phenomena}.  Unbounded-be%
low negative energies that \e{require} ``reinterpretation'' are
a staggeringly bizarre manifestation of \e{undiluted metaphy%
sics} that physical science would do extremely well indeed
to shed forever, \e{especially} in light of the fact that their
\e{origin} resides in gratuitously-generated \e{entirely extran%
eous} ``solutions'' (Klein-Gordon theory) or an equally \e{gratu%
itous weakening} of the correspondence principle that springs
from extraordinary  insistence on, for a physically \e{nonvia%
ble} reason, \e{dealing with the squares of dynamical variables
instead of those dynamical variables themselves} (Dirac theory).
The ``reinterpretation'' of the egregiously metaphysical
unbounded-below negative-energy spectra of the Klein-Gordon and
Dirac theories is a quintessential instance of the syndrome
that afflicted Einstein's knight, ``who dyed his whiskers green,
and then used a large fan so that they should not be seen.''

The \e{breaking} of symmetries has become a very strong theme in
the last seven decades, and the breaking of the antiparticle-%
associated CP invariance has been empirically firmly established.
Indeed a glance at the gross particle-antiparticle ``nonsymmetry''
that makes our galaxy's existence \e{possible} powerfully suggests
that there is \e{much} that remains to be \e{learned} about the
breaking of this symmetry.  A theory that assigns \e{two independent
fields} to the description of the particle and its distinguishable
antiparticle makes it \e{far more straightforward} to model what
are no doubt extremely important effects pertaining to that symmetry
breaking.  With two \e{independent} fields one can essay such basics
as a \e{slight mass difference} between particle and antiparticle,
which is certainly \e{not available} if particle and antiparticle
are construed as merely somehow reflecting \e{different parts} of
the \e{same operator's energy spectrum}.

\subsection*{Conclusion}

It has been made very clear in the above that Schr\"{o}dinger's simple
postulates regarding the solitary particle's wave function specify that
the generators of that wave function's infinitesimal space and time
translations are, respectively, the canonical three-momentum operator
and the Hamiltonian operator, and that this fact provides an extremely
satisfactory basis, which is \e{completely compatible with the strong
correspondence principle}, for the \e{fully relativistic version of
solitary-particle quantum mechanics}.  From the standpoint of \e{system%
atic, conservative} physical theory, the utterly clear implication of
the above observations, in light of the strong correspondence principle,
is that the correct Hamiltonian operator for the solitary relativistic
free particle is given by the square-root operator of Eq.~(3), and that
the interaction of a spinless nonzero-mass solitary charged relativistic
particle with an external electromagnetic potential in the context of
quantum mechanics must be described by the quantization of the Hamilton%
ian that corresponds to the fully relativistic version of H.\ A.\ Lo%
rentz' electromagnetic force law, specifically the Hamiltonian of
Eq.~(10f).

Unfortunately, the pellucid implications for \e{the relativistic quantum
mechanics of the solitary particle} of Schr\"{o}dinger's postulates for
the wave function \e{never dawned} on Schr\"{o}dinger himself, nor on Klein,
Gordon, nor Dirac.  That this is the case is made \e{painfully clear} by
Dirac's simultaneously superfluous and extremely damaging imposition of
``space-time coordinate symmetry'' on Schr\"{o}dinger's \e{time-dependent}
wave function equation (Eq.~(1b)), notwithstanding that \e{precisely this
particular appearance of space-time coordinate symmetry} is the \e{most
striking feature} of Schr\"{o}dinger's \e{four-vector} wave function equa%
tion (Eq.~(1c))!  Although these pioneers failed to appreciate even the
\e{very existence} of Eq.~(1c), \e{notwithstanding} that it is a \e{mere
summary} of Schr\"{o}dinger's wave-function postulates, they \e{did} have
an appreciation of the correspondence principle, albeit they regarded it
as a rather  more \e{plastic} concept than the definitive \e{strong form}
which the \e{facts} of both the Hamiltonian phase-space path integral%
~\ct{Ka3} and the \e{completed} version of Dirac's canonical commutation
rule~\ct{Ka4} \e{very clearly reveal}.  But, extremely unfortunately, be%
cause of perceived issues of calculational convenience, these pioneers
gave the correspondence principle short shrift indeed, and instead chose
ostensible ``relativistic quantum mechanics'' routes that \e{terribly dis%
tort} the quantum mechanical implications and impact of the underlying
\e{classical special relativistic mechanics} of the solitary particle
which had been so ably developed and expounded by Lorentz; the \e{gratui%
tous injection} of \e{totally extraneous} unphysical, unbounded-below
negative-energy ``solutions'' is, of course, a prime example of this.
For any half-way serious disciple of the correspondence principle, these
\e{patently absurd} ``solutions'', which are at \e{complete loggerheads}
with the \e{classical} relativistic solitary-particle mechanics of Lo%
rentz, would \e{of themselves} have been \e{reason enough to immediately
call off} the attemped tweaking of the time-dependent Schr\"{o}dinger
equation and/or of the ostensibly ``relativistic'' Hamiltonians being fed
to it, and to \e{return forthwith} to \e{strictly Lorentzian classical
relativistic basics} as the \e{correct physical foundation} upon which to
erect relativistic quantum mechanics.

Notwithstanding subsequent ``reinterpretation'' of these unbounded-below
negative-energy ``solutions'' to ``accommodate'' the physical existence
of mass-degenerate antiparticles, which is universally acclaimed as a
``triumph''~\ct{B-D}, it is \e{vastly less conceptually tortuous} and
equally vastly \e{more in keeping with all of the rest of known quantum
physics} to account for the existence of mass-degenerate antiparticles
as the \e{utterly straightforward consequence} of \e{CP invariance of
the overlying quantum field theory}.  Since this invariance is, indeed,
\e{a slightly broken one} in the manner of so many \e{other} symmetries
of physics---a fact that would appear to be of critical importance to the
\e{very existence} of the familiar physical world of our experience---it
would seem all the more important to \e{have available} the theoretical
tool of \e{two completely independent quantum fields} (with each having
\e{bounded-below energy}, of course) for the descriptions, respectively,
of a particle and its distinguishable antiparticle, which permits their
masses, for example, to be very slightly different, \e{instead} of being
\e{tied} to the ``reinterpreted negative energy spectrum'' \e{consequence}
of having certain attributes of particle and antiparticle being \e{irrevo%
cably identical}.  It is \e{symmetries} and \e{not} ``reinterpretations''
of blatantly gratuitous and absurdly unphysical equation ``solutions'' that
are reasonably and plausibly held to be responsible for the existence of
mass-degenerate partner particles in \e{all instances other than that of
distinguishable antiparticles}.  It is surely now \e{well past time} for
the theoretical physics treatment of mass-degenerate distinguishable anti%
particles to be put on a track which is \e{completely parallel} to the sen%
sible symmetry-based handling accorded \e{as a matter of standard routine}
to \e{all other} ostensibly ``understood'' particle mass degeneracies.

Departure from the theoretical physics scene of the physically misbehav%
ing Klein-Gordon and Dirac equations would \e{not only} restore physical
good sense to relativistic solitary-particle quantum mechanics, it would
unveil physics' harmonious hierarchical integrity: the \e{underlying phy%
sical essence} of the relativistic Lorentzian Hamiltonian of Eq.~(10f) is
\e{explicitly} its \e{simple nonrelativistic counterpart} of Eq.~(10a),
and the \e{same role} is played by the nonrelativistic Pauli theory of
Eq.~(11a) in relation to \e{its} fully relativistic upgrade that is given
by Eqs.~(11g) and (11f).  Likewise, when the upgrade of these solitary-%
particle theories to fully interacting field theories is made via second
quantization and the imposition of the dual symmetries of CP invariance
and ``CP-equivalence'', the physical essence of the \e{multi-particle in%
teractions} is obviously \e{firmly anchored} in those of the \e{merely
solitary relativistic particle with external fields}, which interactions,
in turn, as we have \e{just noted}, may be no more than the \e{relativis%
tic upgrade} of such \e{nonrelativistic basics} as a stationary charged
particle's interaction with an electric potential or a stationary magne%
tic dipole's interaction with a magnetic field!  Likewise, the properties
of any quantized Hamiltonian are highly sensitive to those of its \e{clas%
sical precursor}, as the key \e{stationary phase path for the path inte%
gral} amply attests.  And of course both the presence \e{and absence} of
symmetries is of crucial importance to the \e{very character} of the phys%
ics.  It is \e{precisely because} of the \e{profound underlying linkages}
inherent in what at first glance may appear to be \e{very diverse} aspects
of physics that the fatuous mathematical ``modifications'' unwarily intro%
duced into relativistic quantum mechanics by Klein, Gordon, Schr\"{o}din%
ger, and Dirac are so very damaging.


\begin{thebibliography}{12}
\bibitem{Scf}
L. I. Schiff,
\e{Quantum Mechanics}
(McGraw-Hill, New York, 1955).
\bibitem{B-D}
J. D. Bjorken and S. D. Drell,
\e{Relativistic Quantum Mechanics}
(McGraw-Hill, New York, 1964).
\bibitem{Dir}
P. A. M. Dirac,
\e{The Principles of Quantum Mechanics}
(Oxford University Press, London, 1958).
\bibitem{Ka1}
S. K. Kauffmann,
arXiv:0909.4025 [physics.gen-ph]
(2009).
\bibitem{Ka2}
S. K. Kauffmann,
arXiv:1004.1820 [physics.gen-ph]
(2010).
\bibitem{D28}
P. A. M. Dirac,
Proc.\ Roy.\ Soc.\ (London) \textbf{A117},
610 (1928).
\bibitem{Scw}
S. S. Schweber,
\e{An Introduction to Relativistic Quantum Field Theory}
(Harper \& Row, New York, 1961).
\bibitem{B-S}
N. N. Bogoliubov and D. V. Shirkov,
\e{Introduction to the Theory of Quantized Fields}
(Interscience Publishers, New York, 1959).
\bibitem{Ka3}
S. K. Kauffmann,
arXiv:0910.2490 [physics.gen-ph]
(2009).
\bibitem{Ka4}
S. K. Kauffmann,
arXiv:0908.3755 [quant-ph]
(2009).
\bibitem{B-J}
M. Born and P. Jordan,
Z.\ Physik \textbf{34},
858 (1925).
\bibitem{L-L}
L. D. Landau and E. M. Lifshitz,
\e{The Classical Theory of Fields}
(Butterworth-Heinemann, Oxford, 1975).
\bibitem{Ha}
G. E. Hahne,
private correspondence.
\end{thebibliography}
\end{document}